\documentclass[english,aps,amssymb,prb, twocolumn,showpacs]{revtex4-1}
\usepackage[T1]{fontenc}
\usepackage[latin9]{inputenc}
\usepackage{amsmath}
\usepackage{graphicx}
\usepackage{amssymb}
\usepackage{esint}

\makeatletter
\renewcommand{\phi}{\varphi}
\renewcommand{\epsilon}{\varepsilon}

\renewcommand{\vec}[1]{{\bf #1}}
\newcommand{Û}{$}
\usepackage{epsfig}%\usepackage{showlabels}
\@ifundefined{showcaptionsetup}{}{%
\PassOptionsToPackage{caption=false}{subfig}}
\usepackage{subfig}
\makeatother

\usepackage{babel}

\begin{document}

\title{Anomalous electromagnetic response of superconducting Rashba systems in trivial and topological phases}

\author{Teemu Ojanen$^{1,2}$}
\email[Correspondence to ]{teemuo@boojum.hut.fi}
\author{Takuya Kitagawa$^2$}
\affiliation{$^1$Low Temperature Laboratory, Aalto University, P.~O.~Box 15100,
FI-00076 AALTO, Finland }
\affiliation{$^2$Physics Department, Harvard University, Cambridge, Massachusetts 02138, USA}

\date{\today}
\begin{abstract}
Two-dimensional electron systems with spin-orbit coupling in the proximity of a superconductor and a magnetic insulator have recently been considered as promising candidates to realize topological superconducting phases. Here we discuss electromagnetic response properties of these systems. Breaking of time-reversal symmetry leads to an anomalous Hall effect with a characteristic non-monotonic gate voltage dependence and a Hall conductivity that can change a sign as temperature is varied. The imaginary part of the Hall conductivity at finite frequency, which shows up for example in the Kerr rotation or photoabsorption, can distinguish different topological
phases. In addition, we demonstrate the existence of magnetoelectric effects associated with the Hall effect;
in-plane electric fields induce a parallel magnetization and in-plane time-dependent magnetic fields induce parallel electric current.

\end{abstract}
\maketitle
\bigskip{}
\section{Introduction}
A spin-orbit coupling of electrons in solid-state systems provides an important link between spin and orbital degrees of freedom, resulting in a wide variety of
phenomena that enable spintronics applications and realizations of novel phases of matter. It has been envisioned that spin-active components could serve as building blocks of future electronic components. \cite{aws} Electrical manipulation of spin, enabled by the spin-orbit coupling, plays a central role in these developments.
Moreover, spin-orbit effects are a crucial ingredient of recently discovered topological insulator materials.\cite{kane2, schnyder} New phenomena and potential for applications have attracted enormous interest in topological insulator physics.

Interesting topological properties exist not only in insulators but also in so-called topological superconductors (TS).
Perhaps the most interesting property of TS is the existence of Majorana fermions which could serve as a platform for topological quantum computation.
One quantum bit can be encoded in two localized Majorana zero modes and computation operations can be carried out by braiding such objects.
Recently it was proposed that topological superconductors can be realized in spin-orbit coupled electron systems in the proximity of a superconductor and a ferromagnetic insulator or in the presence of magnetic fields.\cite{lutchyn1,sato,lutchyn,oreg,alicea}
Besides the interests in TS, the interplay of superconductivity, a spin-orbit coupling and magnetization is interesting in its own right, since there already exists a number of experimental realizations coupling a 2DEG with a superconductor.\cite{deon}

In this paper we study unusual electromagnetic properties of a Rashba-coupled superconducting two-dimensional electron gas (2DEG) in the presence of magnetization perpendicular to the plane. One the most important phenomenon resulting from the Rashba coupling and time-reversal symmetry breaking due to the magnetization is the anomalous Hall effect.\cite{nagaosa} Previously the anomalous Hall effect in a spin-orbit coupled superconducting system
has been studied only numerically in a lattice model with a single magnetic impurity.\cite{sacramento} Here we find that the studied system exhibits a characteristic anomalous Hall effect which is non-monotonic as a function of chemical potential. The Hall conductivity can even change sign when temperature or chemical potential are varied. In stark contrast to translationally invariant chiral p-wave systems which also break time-reversal symmetry,\cite{lutchyn2,lutchyn3,roy,stone} the Hall conductivity of the studied system remain finite in the low frequency limit.
Moreover, the onset frequency of the dissipative part of the ac Hall conductivity behaves qualitatively differently in topologically trivial and non-trivial phases, enabling an electrical characterization of phases. Topological nature of TS manifests in the quantized thermal conductivity, \cite{read} measurement of which is challenging in experiments. The electromagnetic response studied here provides a signature of topological phase transition which should be easier to measure. Intriguingly, there exists also magnetoelectric responses intimately related to the Hall effect; in-plane electric fields induce parallel magnetization and in-plane time-dependent magnetic fields induce parallel electric currents. These magnetoelectric effects are unique properties of Rashba systems in the presence of magnetization and have no counterparts in chiral p-wave systems.

\section{Model and electromagnetic action}
\begin{figure}[t]
\centering
\includegraphics[width=0.9\columnwidth, clip=true]{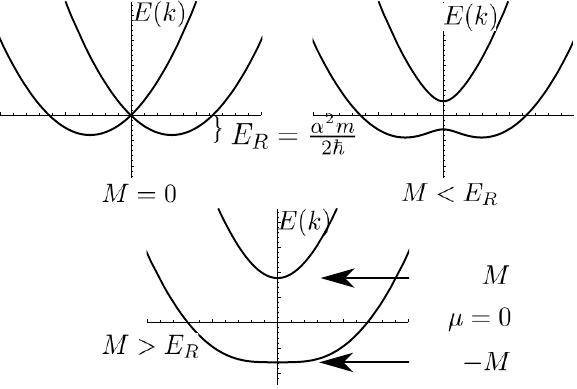}
\caption{  Normal state spectrum as function of the in-plane momentum ÛkÛ for different values of perpendicular magnetization  $M$.
The strength of the Rashba coupling is characterized by $E_{R} = \alpha^2 m/2\hbar$, chemical potential is measured from the middle of the magnetization induced gap.}
\label{spect}
\end{figure}

In this paper we study a 2DEG with a Rashba spin-orbit coupling and magnetization in the proximity of a s-wave superconductor.
The system is described by a Bogoliubov-de Gennes Hamiltonian\cite{lutchyn1},
\begin{align}\label{h}
H(\vec{k},\phi)=&\left(\epsilon_k+\alpha(k_x\sigma_y-k_y\sigma_x)\right)\tau_z+\nonumber \\ &M\sigma_z+\Delta \mathrm{cos}\,\phi\,\tau_x+\Delta \mathrm{sin}\,\phi\,\tau_y,
\end{align}
where $\epsilon_k=\frac{\hbar^2k^2}{2m}-\mu$ and $\sigma_i$ and $\tau_i$ are Pauli matrices operating in the spin and the particle-hole space, respectively. Hamiltonian (\ref{h}) is written in the Nambu basis  $\Psi=(\psi_{k\uparrow}, \psi_{k\downarrow},\psi^{\dagger}_{-k\downarrow},-\psi^{\dagger}_{-k\uparrow} )^T$. The first term corresponds to the kinetic energy of electrons and holes including the Rashba coupling, the second term is the Zeeman splitting due to out-of-plane magnetization and the last two terms are proximity-induced superconducting pairing terms for the order parameter $\Delta\,e^{i\phi}$. Fundamental properties of the model have been discussed in detail in Refs.~\onlinecite{sau1,alicea1} and the effects of disorder have been considered in Refs.~\onlinecite{lutchyn4, potter} 
In the absence of superconducting order, the energy bands are illustrated in Fig.~\ref{spect} for various magnetizations and spin-orbit energies ÛE_R=\alpha^2 m/2\hbarÛ. We will see below that
qualitative features of the Hall response depend heavily on the location of chemical potential.
The spectrum of (\ref{h}) consists of four bands and is symmetric with respect to zero energy $E=0$ due to
the presence of a particle-hole symmetry. The positive energy bands $E_{1}(\vec{k})$, $E_{2}(\vec{k})$ and the negative energy bands $E_{-1}(\vec{k})$, $E_{-2}(\vec{k})$ satisfy $E_{-i}(\vec{k}) = -E_{i}(\vec{k})$ for $i=1,2$.
The energies are given by
\begin{align}\label{E}
E_{1/2}^2(\vec{k})=\epsilon_k^2&+\alpha^2k^2+M^2+\Delta^2\nonumber \\ &\mp2\sqrt{M^2(\epsilon_k^2+\Delta^2)+\epsilon_k^2\alpha^2k^2}
\end{align}
with $k =|\vec{k}|$, so the spectrum is rotationally symmetric.
\begin{figure}[h]
\centering
\includegraphics[width=0.5\columnwidth, clip=true]{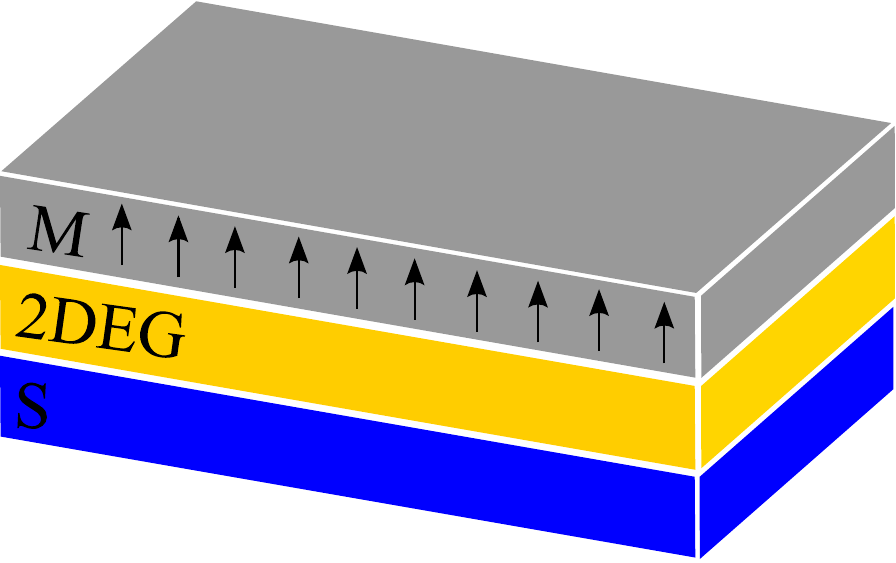}
\caption{ Studied heterostructure formed by  2DEG with a Rashba spin-orbit coupling, sandwiched by a ferromagnetic insulator and a superconducting layer.  The coordinates are chosen so that the induced magnetization, which is perpendicular to the 2DEG, coincides with the ÛzÛ axis and the 2DEG  lies in the Ûx-yÛ plane. }
\label{fig1}
\end{figure}
The effective action involving only electromagnetic fields can be derived by integrating out electronic as well as
the superconducting phase degrees of freedom, as detailed in Ref.~[\onlinecite{lutchyn2}] in the context of a chiral p-wave system.
The advantage of this procedure is that it yields an action which is explicitly gauge-invariant. The starting point is
the action corresponding to Eq. (\ref{h}) which is $S=\int d^2x\,d\tau\Psi^\dagger\left[\partial_\tau-H(k,\phi) \right]\Psi/2$.
The system is coupled to electromagnetic potentials through the substitution $\partial_\tau\to \partial_\tau-ieA_0\tau_z$ and $\vec{k} \to \vec{k}-e\boldsymbol{A}\tau_z/\hbar$, where $A_0$ is the scalar potential and $\boldsymbol{A}=(A_x,A_y)$ is the vector potential. Below we combine the potentials to a single quantity
$A =(A_0,\boldsymbol{A})$. It is convenient to perform a gauge transformation through the unitary rotation $U=e^{i\frac{\phi\,\tau_z}{2}}\Psi$, which leads to
the action
\begin{align}\label{S2}
S&=\frac{1}{2}\int d^2x\,d\tau\nonumber\\
&\Psi^\dagger\left[\partial_{\tau}-ie\tilde{A}_0\tau_z-H(\vec{k}-e\boldsymbol{\tilde{A}}\tau_z/\hbar, 0)-\vec{B}\cdot\vec{\sigma} \right]\Psi,
\end{align}
where $\tilde{A}_0=A_0-\frac{\partial_\tau\phi}{2e}$, $\boldsymbol{\tilde{A}}=\boldsymbol{A}-\frac{\hbar\nabla\phi}{2e}$. We have included the Zeeman coupling due to in-plane magnetic fields with Û\vec{B}=\frac{1}{2}g\mu_B(b_x,b_y)Û, where ÛgÛ is the effective ÛgÛ-factor, Û\mu_BÛ is the Bohr magneton and Û(b_x, b_y)Û the in-plane magnetic field. Integrating out electronic degrees of freedom yields
$\int \mathcal{D}\Psi^\dagger\mathcal{D}\Psi e^{-S}=e^{-S'[\tilde{A},B]}$, where $S'[\tilde{A},B]$ is given by the determinant of the kernel
in Eq. (\ref{S2}). The saddle-point expansion up to the second order in the field variables produces
\begin{align}
S&'[\tilde{A},B]=\frac{1}{2}\sum_q \left[\right. \tilde{A}_\mu(-q)Q^{(1)}_{\mu\nu}(q)\tilde{A}_\nu(q)+\nonumber\\
&\left. B_i(-q)Q^{(2)}_{ij}(q)B_j+2 B_{i}(-q)Q_{i\mu }^{(3)}(q)\tilde{A}_\mu(q) \right]\nonumber, 
\end{align}
where we have adopted a convention that repeated indices should be summed.
Quantities $Q^{(1)}_{\mu\nu}$ are the standard current-current correlation functions describing the response to the dressed electromagnetic fields $\tilde{A}$ which
also contains dynamic of the superconducting phase $\phi$.  The spin-spin and the spin-current correlation functions defined as  $Q^{(2)}_{ij}(q)= \langle \sigma_i\sigma_j \rangle(q)$  and $Q^{(3)}_{i\mu}(q)= \langle \sigma_iJ_{\mu} \rangle(q)$ where the angular brackets stand for two-point functions calculated for vanishing fields $A_\mu=0$, $B_i=0$ and $\phi=0$ and $J_\mu$ denotes the appropriate current operator discussed below. The true response is obtained by further
integrating out $\phi$, leading to $\int \mathcal{D}\phi e^{-S'[\tilde{A},B]}=e^{-S_{\mathrm{eff}}[A,B]}$
where the effective electromagnetic action is
\begin{align}
&S_{\mathrm{eff}}[A,B]=\frac{1}{2}\sum_q \left[\right. A_\mu(-q)K^{(1)}_{\mu\nu}(q)A_\nu(q)+\nonumber\\
&\left. B_i(-q)K^{(2)}_{ij}(q)B_j+2 B_{i}(-q)K_{i \mu }^{(3)}(q)A_\mu(q) \right]\nonumber. 
\end{align}
The kernels  are  given by
\begin{align}\label{ker}
&K_{\mu\nu}^{(1)}(q)=Q_{\mu\nu}^{(1)}(q)-\frac{q_\alpha q_\beta Q_{\mu\alpha}^{(1)}(-q)Q_{\beta\nu}^{(1)}(q)}{q_\alpha q_\beta Q_{\alpha\beta}^{(1)}(q)},\nonumber\\
&K_{ij}^{(2)}(q)=Q_{ij}^{(2)}(q)-\frac{q_\alpha q_\beta Q_{i\alpha}^{(3)}(-q)Q_{\beta j}^{(3)}(q)}{q_\alpha q_\beta Q_{\alpha\beta}^{(1)}(q)},\nonumber\\
&K_{i \mu}^{(3)}(q)=Q_{i \mu }^{(3)}(q)-\frac{q_\alpha q_\beta Q_{i\alpha}^{(1)}(-q)Q_{\beta \mu}^{(3)}(q)}{q_\alpha q_\beta Q_{\alpha\beta}^{(1)}(q)},
\end{align}
where $q=(\omega,q_x,q_y)$. 
The density and current responses are given by $\delta J_\mu(q)=K_{\mu\nu}^{(1)}(q)A_{\nu}(q)+K_{\mu\nu}^{(3)}(q)B_{\nu}(q)$, where $\delta J_\mu=(i\delta\rho,\delta\boldsymbol{J})$.  We have now arrived at purely electromagnetic action which is explicitly current conserving $q_{\mu}\delta J_\mu(q)=0$. 

So far the manipulation has been formal and independent of the detailed form of Eq. (\ref{h}). Physical properties of the electromagnetic response (\ref{ker}) are encoded in the specific form of functions $Q_{\mu\nu}^{(i)}$. Below we are mainly interested in the case where the in-plane Zeeman field is absent and the response is given by $Q_{\mu\nu}^{(1)}$. For superconductors the diagonal responses $Q_{\mu \mu}^{(1)}$
are non-zero in the limit $q \to 0$ and $\omega \to 0$. The spatial components at low frequencies are given by the diamagnetic term $Q_{ii}^{(1)}=\frac{e^2n_s}{m}$, where $n_s$ is the superfluid density. The elements $Q_{ii}^{(1)}$ give rise to the Meissner effect which screens the magnetic field in the bulk. The density-density component $Q_{00}{(1)}$ is given by the density-of-states in the normal state. In the following we are concentrating on effects arising from the off-diagonal spatial components $Q_{ij}^{(1)}$ $(i\neq j)$ which contain signatures from the time-reversal symmetry breaking and are responsible for the anomalous Hall effect. We are interested in the current response in the long-wavelength limit $q\to 0$ at finite $\omega$. According to $(\ref{ker})$, the off-diagonal physical response functions are given by $K_{ij}^{(1)}=Q_{ij}^{(1)}$, $K_{i0}^{(1)}=Q_{ij}^{(1)}q_j/\omega$ and $K_{0i}^{(1)}=Q_{ji}^{(1)}q_j/\omega$ $(i\neq j)$ in this limit.

\section{Off-diagonal response $Q_{xy}^{(1)}$}In this section we calculate the off-diagonal current-current response functions $Q_{ij}^{(1)}$ $(i\neq j)$ in the long-wavelength limit.
The expression for the current operator is $J_i=e(\frac{\hbar k_i}{m}+\frac{\alpha}{\hbar}\epsilon_{ij}\sigma_j)$, where
the antisymmetric tensor $\epsilon_{ij}$ is defined as $\epsilon_{11}=\epsilon_{22}=0$, $\epsilon_{12}=-\epsilon_{21}=1$.
The off-diagonal response function in the imaginary-time representation is
\begin{align}\label{qxy1}
Q_{xy}^{(1)}(i\omega_m)=\frac{1}{2\Omega\beta}\sum_{k,n}\mathrm{Tr}[J_xG(i(\omega_m+\nu_n))J_yG(i\nu_n)].
\end{align}
In the above expression  $G(i\omega_n)=\sum_i\frac{P_i}{i\omega_n-E_i}$ is the $4\times4$  Matsubara Green's function of the Hamiltonian
$H=H(\vec{k}, \phi= 0)$, $\beta$ is the inverse temperature and $\Omega$ is the area of the system. The summation over $i$ is performed over the four energy bands and $P_i$ is a $4\times4$ projection operator to the $E_i$ subspace. The trace is evaluated over the spin and Nambu indices. Inserting the expressions for Green's functions, performing the summation over $n$ and analytical continuation to real frequencies yields
\begin{align}\label{fin1}
iQ_{xy}^{(1)}(\omega)=\frac{i}{2\Omega}\sum_{k,i,j}\frac{\mathrm{Tr}[J_xP_iJ_yP_j]}{\omega-E_i+E_j+i\delta}\left(n_j-n_i\right),
\end{align}
where $n_i$ is a Fermi function at energy $E_i$. The projection operators are given by $P_{\pm 1}=\frac{1}{2}\left(1\pm\frac{H}{E_1}\right)\frac{H^2-E_2^2}{E_1^2-E_2^2}$, and analogously for $P_{\pm 2}$ with indices $1$ and $2$ interchanged. With these results the evaluation of the traces is straightforward but tedious. The evaluation is slightly simplified by noting that only the terms that involve spin operators in $J_i$ contribute to Eq.~(\ref{fin1}). Also, contributions for which $i=-j$ in the sum vanish due to the property $\langle E_i|J_{x/y}|E_{-i} \rangle=0$. Collecting all the non-vanishing terms and converting the $k$-summation to integration leads to expression
\begin{align}\label{fin2}
&\frac{iQ_{xy}^{(1)}(\omega)}{\omega}=\frac{e^2}{h}4\alpha^2M\int dk k \frac{1}{E_1E_2} \left[\right.\nonumber\\ &\left. \left(-\frac{\Delta^2+\epsilon_k^2}{E_1+E_2}+\frac{E_1+E_2}{4} \right) \frac{\left(n_1+n_2-1\right)}{(E_1+E_2)^2-(\omega+i\delta)^2}+ \right.\nonumber\\
&\left.\left(\frac{\Delta^2+\epsilon_k^2}{E_2-E_1}-\frac{E_2-E_1}{4} \right) \frac{\left(n_1-n_2 \right) }{(E_2-E_1)^2-(\omega+i\delta)^2} \right].
\end{align}
The expression Eq. (\ref{fin2}) is proportional to $M$ and $\alpha^2$ indicating that both magnetization and the spin-orbit coupling are crucial for the off-diagonal term.
This also implies that the Hall effect is independent of the sign of spin-orbit coupling $\alpha$. This is expected since spatial inversion takes $\alpha \rightarrow -\alpha$
but it does not change the sign of the Hall conductivity.
The off-diagonal functions satisfy $Q_{xy}^{(1)}(\omega)=-Q_{yx}^{(1)}(\omega)$ which is a signature of the Hall effect.
\begin{figure}[t]
\centering
    %\hline
\includegraphics[height=0.4\columnwidth]{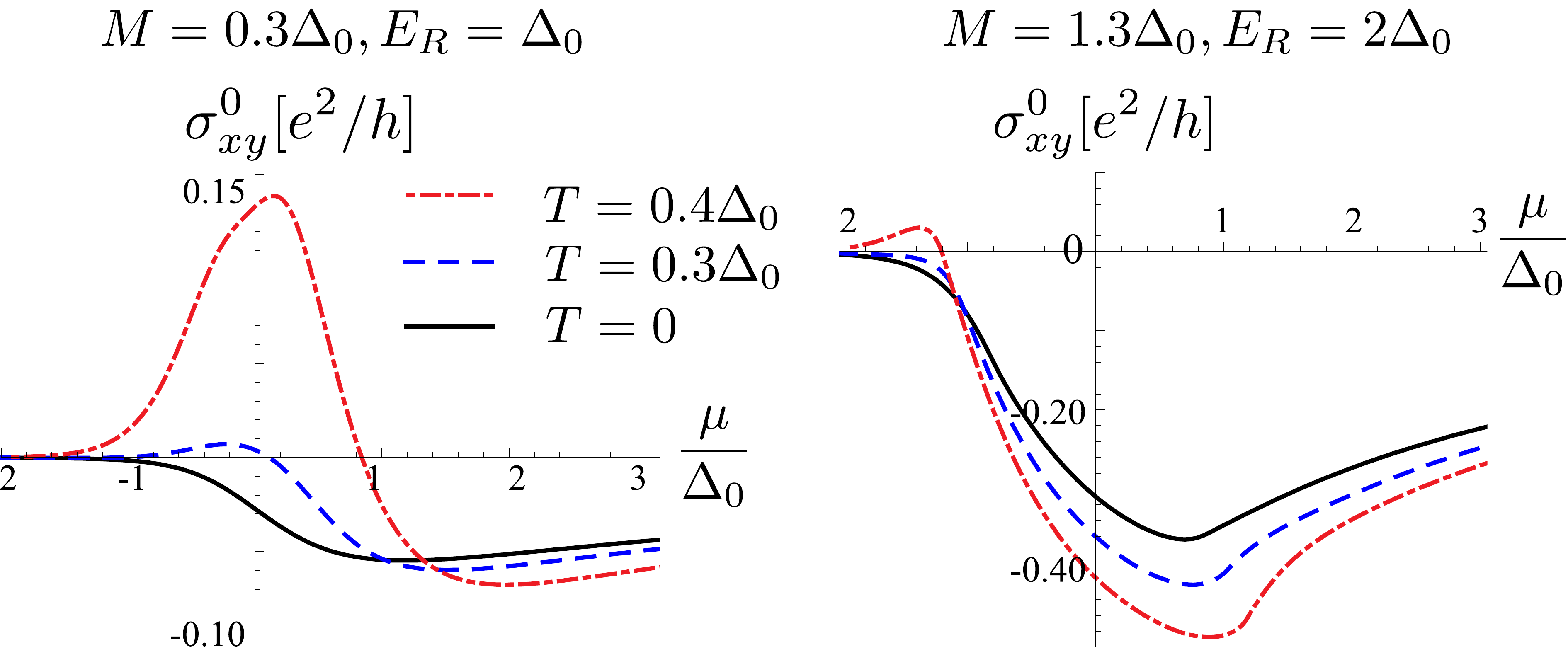}
\caption{ (a):  Low-frequency Hall conductivity as a function of the chemical potential. The temperature dependence of $\Delta$ is assumed to be of the BCS form $\Delta(T)/\Delta_0=\sqrt{1-\frac{T}{T_c}}$. (b): Same quantities as in (a) but for different parameters.}   \label{g1}
\end{figure}

\section{Anomalous Hall effect}
Here we consider properties of the off-diagonal current response to spatially uniform, slowly varying electric fields. As discussed above, the off diagonal response $K_{ij}$  is given by $Q_{ij}^{(1)}$ ($i\neq j$)
%$K_{i0}^\mathrm{a}=\epsilon_{ij}q_jQ_{xy}^\mathrm{a}(\omega)/\omega$
which leads to the Hall effect  $\delta J_i(\omega)=-\frac{iQ_{xy}^{(1)}(\omega)}{\omega}\epsilon_{ij}E_j(\omega)$.
For frequencies $\omega\ll \min_k (E_2(k)-E_1(k))$,  the Hall conductivity is given by $\sigma_{xy}^0=\lim_{\omega\to 0} Q_{xy}^{(1)}(\omega)/i\omega$.
Quantity $\sigma_{xy}^0$ in plotted in Figs.~\ref{g1} (a)-(b) as a function of chemical potential at different temperatures.
In experiments, chemical potential could be varied through a gate voltage. For simplicity we have assumed that the proximity effect is nearly perfect and the induced superconducting gap obeys a standard BCS temperature relation  $\Delta(T)/\Delta_0=\sqrt{1-\frac{T}{T_c}}$, where Û\Delta_0Û is the gap at zero temperature and ÛT_cÛ satisfies Û\Delta_0=1.76k_BT_cÛ.

The intuition of the behavior of $\sigma_{xy}^0$ can be obtained by considering the spectrum in the absence of superconductivity plotted in Fig.~\ref{spect}.
When chemical potential is below $\sim-E_R$, the Hall conductivity is suppressed since the electron density vanishes.
As in the normal state systems,\cite{nagaosa} the band curvature effects giving rise to the Hall conductivity reach maximum for chemical potentials located between the two bands in Fig.~\ref{spect}. For larger chemical potentials the absolute value of the Hall conductivity decreases monotonically. The Hall conductivity can change significantly even when temperature and chemical potential variations are small compared to the energy scale of superconductivity $\Delta$ since magnetization competes with supeconductivity and suppresses the excitation gap. Interestingly, the sign of the Hall conductivity can change as a function of chemical potential and temperature due to the different contributions of band curvatures of the four bands.

The result (\ref{fin2}) reduces in the limit $\Delta\to 0$ and $\omega\to 0$ to the well-known expression of the Berry-curvature contribution to the anomalous Hall conductivity of the magnetic Rashba model. To see this, first consider the first term inside the integral. In the   $\Delta\to 0$ limit the energy bands are given by $E_{1/2}=||\epsilon_k|\mp\sqrt{M^2+\alpha^2k^2}|$. Using this result one can show that
\begin{align}\label{fin3}
&\sigma_1=\frac{e^2}{h}4\alpha^2M\int dk k \frac{1}{E_1E_2} \nonumber\\ &\left. \left(-\frac{\Delta^2+\epsilon_k^2}{E_1+E_2}+\frac{E_1+E_2}{4} \right) \frac{\left(n_1+n_2-1\right)}{(E_1+E_2)^2}=\right.\nonumber\\
&\frac{e^2}{2h}\alpha^2M\int_{\Omega_1} dk k  \frac{\left(n_1+n_2-1\right)}{(M^2+\alpha^2k^2)^{3/2}},
\end{align}
where $\Omega_1$ denotes part in the $k$ space which satisfy $|\epsilon_k|<\sqrt{M^2+\alpha^2k^2}$. Using the property $n(x)+n(-x)=1$ of the Fermi function we can write $n(E_2)+n(E_1)-1=n(E_2)-n(-E_1)$ which further reduces to  $n(E_2)-n(-E_1)=n(|\epsilon_k|+\sqrt{M^2+\alpha^2k^2})-n(-|\epsilon_k|+\sqrt{M^2+\alpha^2k^2})$  in the studied parameter regime. Finally, using the property $n(x_1)-n(x_2)=n(-x_2)-n(-x_1)$, we can drop the absolute value signs from $|\epsilon_k|$ and write $n(E_2)+n(E_1)-1=n(\epsilon_2)-n(\epsilon_1)$, where $\epsilon_{1/2}=\epsilon_k\mp\sqrt{M^2+\alpha^2k^2}$. Thus we obtain
\begin{align}\label{fin4}
&\sigma_1=\frac{e^2}{2h}\int_{\Omega_1} dk k  \frac{\alpha^2M}{(M^2+\alpha^2k^2)^{3/2}}\left[n(\epsilon_2)-n(\epsilon_1)\right].
\end{align}
Similarly one can show that the second term inside the integral of Eq.~(\ref{fin2}) reduces to (\ref{fin4}) in the complementary region in the $k$ space where $|\epsilon_k|>\sqrt{M^2+\alpha^2k^2}$. Adding both contributions together lead to 
\begin{align}\label{fin5}
\sigma_{xy}=\frac{e^2}{\hbar}\int \frac{d^2\mathbf{k}}{8\pi^2} \frac{\alpha^2M}{(M^2+\alpha^2k^2)^{3/2}}\left[n(\epsilon_2)-n(\epsilon_1)\right],
\end{align}
which exactly coincides with the previously known expression\cite{nagaosa} arising from the Berry curvature in a clean system. Thus we see that our general expression for Hall conductivity is in perfect agreement with the known expression of the normal system. The Hall conductivity is a continuous function through the superconducting transition.

The system undergoes a topological phase transition between a trivial and a topological phase when $\mu$, $\Delta$ and $M$ satisfy the condition $\mu^2+\Delta^2=M^2$. \cite{lutchyn1} Unfortunately, $\sigma_{xy}^{0}$ is not sensitive to the phase transition. The phase transition is accompanied by a closing of the energy gap in the center of the Brillouin zone such that $E_1(0)=0$. Vanishing denominator in Eq. (\ref{fin2}) at the transition point is compensated by the vanishing numerator, so $\sigma_{xy}^0$ is a smooth continuous function at the phase transition point. Even though the parameters corresponding to Fig.~\ref{g1} (b) are chosen so that the system undergoes a phase transition at finite chemical potential, the precise point of the transition is not visible. However, we show below that the dissipative part of the Hall conductivity at finite frequency can clearly detect the phase transitions.

\section{Identifying topological phases}
\begin{figure}[t]
\centering
\begin{tabular}{c c } %{|c||c||c|}
    %\hline
%\label{a} \includegraphics[width=0.60\columnwidth, clip=true]{Freqzero3}
\includegraphics[height=0.32\columnwidth]{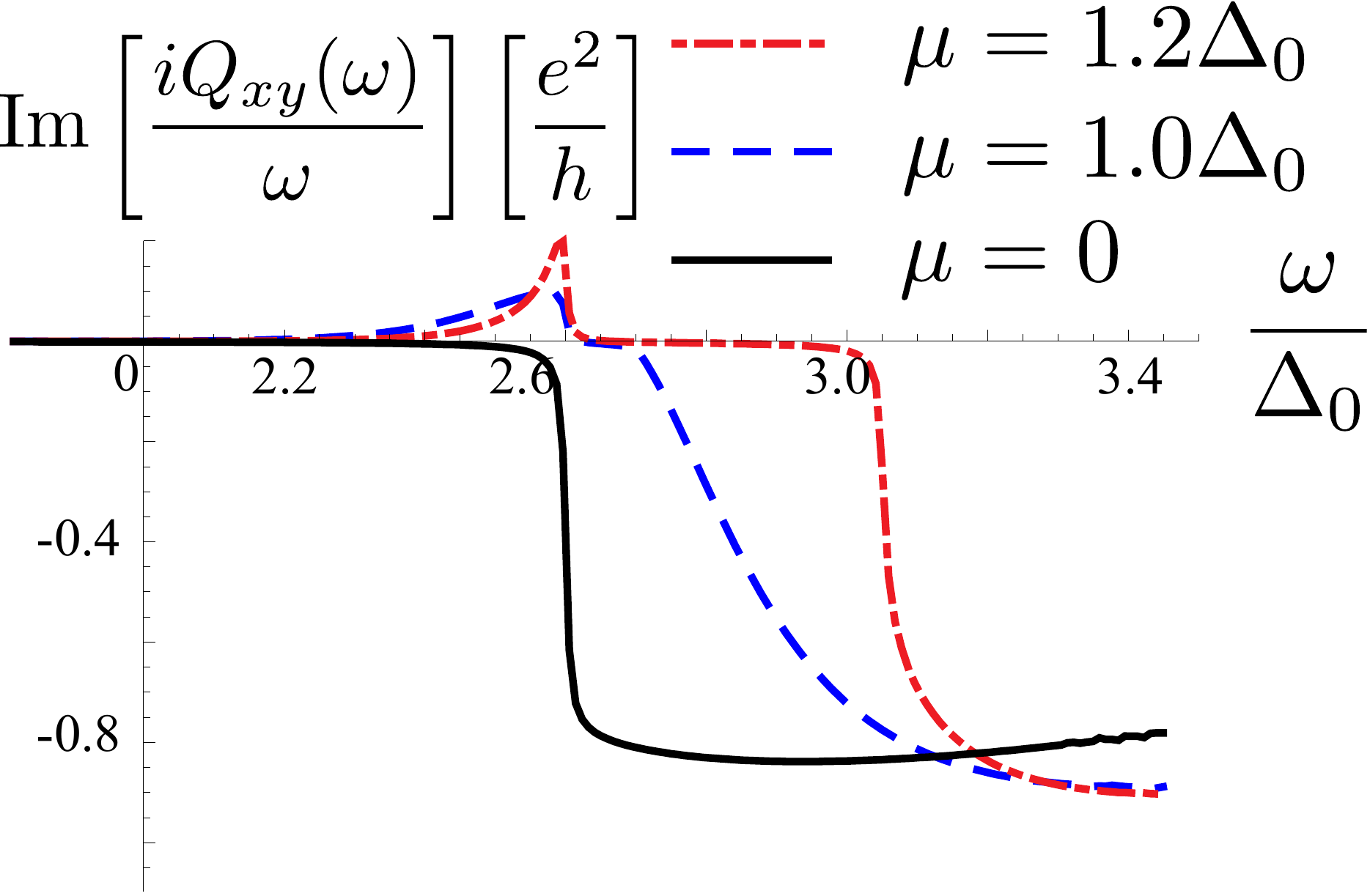}& %
\includegraphics[height=0.27\columnwidth]{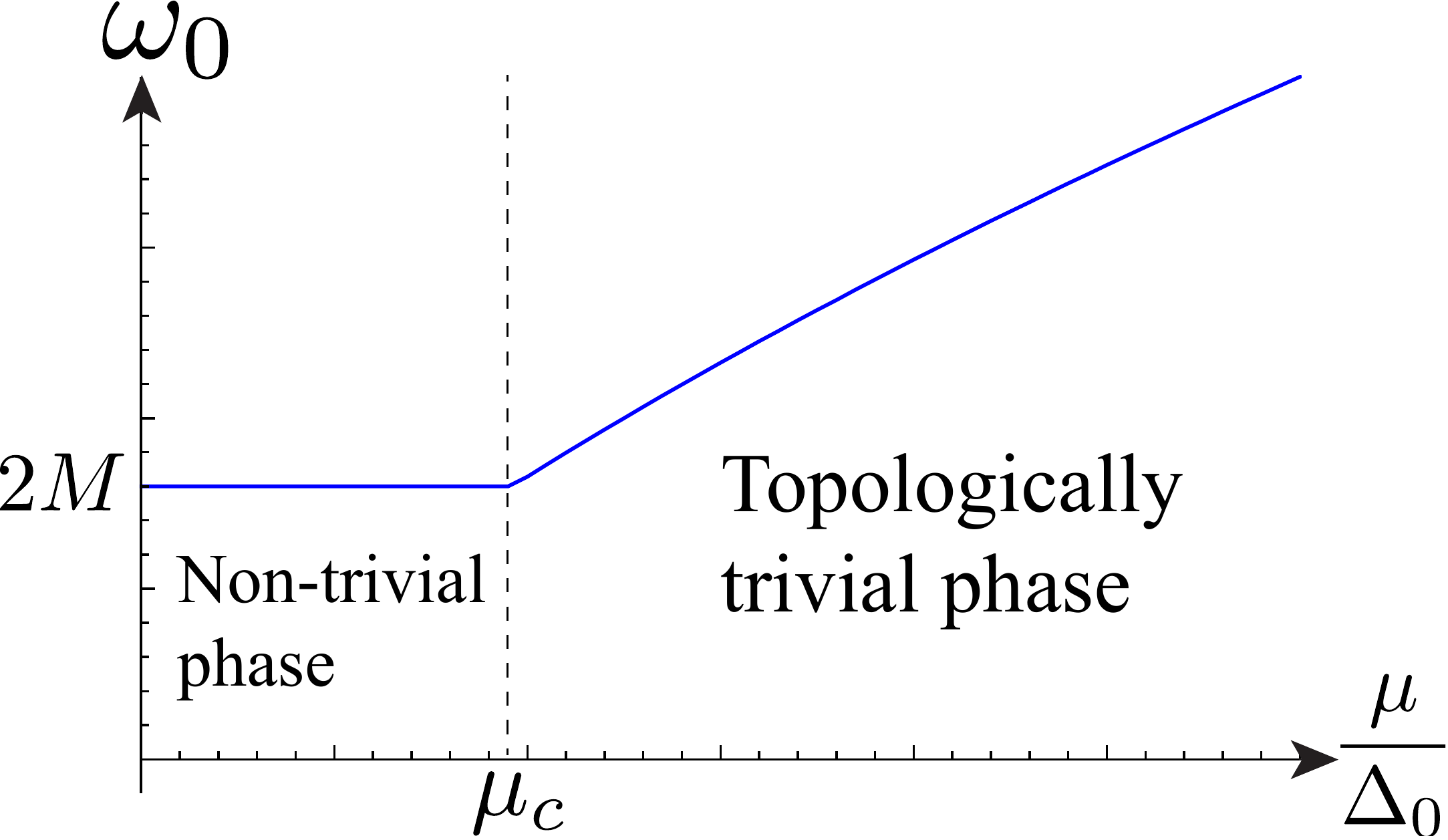} \\
(a)&(b) %\hline
% \vspace{-5mm}
 \end{tabular}
\caption {(a): Imaginary part of the Hall conductivity as a function of frequency at $T=0.1\Delta_0$, where Û\Delta_0Û is the gap at ÛT=0Û. For all curves $M=1.3\Delta_0$, $E_R=2\Delta_0$. The threshold frequency for the nontrivial phase (solid line) is $\omega_0=2M$ above which the imaginary part is finite at $T=0$. Dashed lines correspond to trivial phases. Thermal excitations show up in the positive peaks that are more pronounced for trivial phases. (b): Zero temperature threshold frequency as a function of chemical potential. Existence of a plateau signals a nontrivial phase.}   \label{g2}
\end{figure}

At zero temperature the contribution on the third line of Eq.~(\ref{fin2}) vanishes, so the imaginary part of the Hall conductivity is finite only for frequencies $\omega>\omega_0=\min_k (E_1(k)+E_2(k))$. In the topologically nontrivial phase where $M^2-\Delta^2-\mu^2>0$, the minimum always takes place at $k=0$ and the threshold frequency is $\omega_0=2|M|$ which is \emph{independent} of the chemical potential. In contrast, in the trivial phase $M^2-\Delta^2-\mu^2<0$ the threshold frequency $\omega_0$ is always increasing (decreasing) function of $\mu$ for $\mu>0$ ($\mu<0$). In the trivial phase for small chemical potentials $|\mu| \lesssim \Delta$ the minimum $\min_k (E_1(k)+E_2(k))$ also takes place at $k=0$ but the threshold frequency is $\omega_0=2\sqrt{\Delta^2+\mu^2}$. Thus the nontrivial phase is characterized by horizontal plateaus in the plot $\omega_0$ vs. $\mu$, terminating at critical values $\mu_c=\pm\sqrt{M^2-\Delta^2}$ corresponding to phase transitions to the trivial phase, as illustrated in Fig.~\ref{g2}. By measuring the onset frequency $\omega_0$ while varying chemical potential, \emph{it is possible determine whether the system is in the topological or in the trivial phase}.

Experimentally it has been verified that the proximity induced gap in 2DEG may be at least of the order of Û\Delta_0\sim 0.1Û meV.\cite{deon}  Assuming that ÛMÛ can be made also be of that order (and somewhat larger to enter the topological phase), the threshold frequency Û\omega_0Û is of the order of few tens of GHz. This provides the upper limit of the frequencies of the interest in the identification of the topological phases.        

Characterization of TS phases by electrical means like proposed here or in tunneling experiments proposed in Ref.~\cite{Yamakage} is important since measurement of quantized thermal conductivity, the natural topological invariant of TS, is very difficult. Both the real and imaginary part of the Hall conductivity can be
probed optically through the Kerr effect, which has been employed to
characterize broken time-reversal symmetry in the p-wave candidate Sr$_2$RuO$_4$.\cite{xia, mineev} This technique requires comparing polarizations of incident and reflected electromagnetic waves.

\section{Magnetoelectric effects} \label{last}
An interesting consequence of the non-zero Hall conductivity is the existence of unusual magnetoelectric effects.
First we consider the response of the system to in-plane electric fields $E_i(\omega)=i\omega A_i(\omega)$, which results in magnetization \emph{parallel} to the applied fields. To derive this effect, we consider induced magnetization $\rho_{S_i}=\frac{\hbar}{2}\frac{\langle \sigma_i\rangle}{\Omega}$ which  can be expressed in  the standard linear-response theory as 
\begin{align}\label{mag0}
\rho_{S_i}(\omega)=\frac{\hbar}{2}K_{ij}^{(3)}(\omega)A_j(\omega),
\end{align}   
where $K_{ij}^{(3)}(\omega)$ is the spin-current response function (\ref{ker}).  In the long wavelength limit $\mathbf{q}\to 0$ at finite $\omega$ the response function reduces to  $K_{ij}^{(3)}(\omega)=Q_{ij}^{(3)}(\omega)$  with the imaginary-time representation given by
\begin{align}\label{chi1}
Q_{ij}^{(3)}(i\omega_m)=\frac{1}{2\Omega\beta}\sum_{k,n}\mathrm{Tr}[\sigma_iG(i(\omega_m+\nu_n))J_jG(i\nu_n)].
\end{align}
Since the  off-diagonal current response (\ref{qxy1}) arises solely from the spin part of the current operator, we see that the diagonal component of expression (\ref{chi1}) is give by $Q_{yy}^{(3)}(\omega)=\frac{\hbar Q_{xy}^{(1)}(\omega)}{e\alpha}$. Therefore the magnetization parallel to the electric field is related to the Hall conductivity as
\begin{align}\label{mag}
\rho_{S_y}(\omega)=\frac{\hbar^2}{2e\alpha} \frac{iQ_{xy}^{(1)}(\omega)}{\omega} E_y(\omega)=\frac{\hbar^2}{2e\alpha}\sigma_{xy}^0 E_y(\omega),
\end{align}
where we took the low-frequency limit. Since the system is rotationally invariant, similar relation holds also in the $x$ direction.
This effect can be intuitively understood as follows. The application of electric field induces Hall current in the perpendicular direction.
Because of Rashba coupling, current is accompanied by magnetization perpendicular to current.\cite{edelstein} Thus, the application of an electric field
results in magnetization parallel to it. This phenomenon is a consequence of the Rashba coupling and a finite out-of-plane magnetization $M\neq 0$. No analogous effect exists in chiral p-wave superconductors. The Rashba coupling also results in a previously discovered magnetization \emph{perpendicular} to applied electric fields which remains finite also for vanishing out-of-plane magnetization $M=0$.\cite{edelstein}

There exists also a Zeeman-type magnetoelectric effect closely related to the Hall effect.
Suppose that the system is exposed to an in-plane magnetic field $B_x$ which couples to the spin of the particles through a Zeeman term $ B_x\sigma_x$.
Similarly as before, we can analyse the linear response of the current  to the Zeeman field. Analogously we find 
\begin{align}\label{mag1}
\delta J_i(\omega)=K_{ij}^{(3)}(\omega)B_j(\omega),
\end{align}
Therefore, for similar reasons than discussed above, the relevant response function can be expressed in the long wavelength limit as $K_{xx}^{(3)}(\omega)=Q_{xx}^{(3)}(\omega)=-\frac{\hbar Q_{xy}^{(1)}(\omega)}{e\alpha}$. Thus the parallel part of the current response to the applied Zeeman field is given by
\begin{align}\label{mag1}
\delta J_x(\omega)=-\frac{\hbar}{e\alpha}\frac{Q_{xy}^{(1)}(\omega)}{i\omega}i\omega B_x(\omega)=-\frac{\hbar}{e\alpha}\sigma_{xy}^0 \partial_tB_x(\omega),
\end{align}
where the last form is valid for low frequencies. The magnitude of the effect depends on the Hall conductivity and the ratio of the magnetic moment and the spin-orbit constant $\alpha$.
Interestingly, this magnetoelectric response flips the sign upon the changing of the sign of $\alpha$, in contrast to the Hall effect which is insensitive to the sign. It should be noted that the parallel current (\ref{mag1}) can be distinguished from the field induced Meissner current and the previously studied paramagnetic current\cite{edelstein}  by its different directional and functional dependence on the magnetic field. For a linearly polarized in-plane magnetic field current (\ref{mag1}) is parallel (or antiparallel) to the applied field, not perpendicular to it like the other contributions. In further contrast, for static magnetic fields current (\ref{mag1}) vanish. 

\section{Conclusions}We studied electromagnetic response properties of superconducting two-dimensional electron systems with Rashba spin-orbit coupling and perpendicular magnetization. We focused on the anomalous Hall effect and closely related magnetoelectric effects. The anomalous Hall effect is finite in the low-frequency limit, exhibiting a non-monotonic behavior as a function of chemical potential. Moreover, the frequency-dependent Hall conductivity enables a purely electrical characterization of different topological phases. We found two magnetoelectric effects directly related to the Hall effect, one leading to a parallel magnetization as a response to in-plane electric fields and the other giving rise to electric currents parallel to time-dependent in-plane magnetic fields. 

One of the authors (T.O.) would like to thank Academy of Finland for support and T.K. acknowledges support from Army Research Office with funding from the
DARPA OLE program, Harvard-MIT CUA, NSF Grant No. DMR-07-05472,
AFOSR Quantum Simulation MURI, the ARO-MURI on Atomtronics.

\end{document}